\documentclass[aps,prb,twocolumn,floats,showpacs,amsmath,amssymb,floatfix,%
superscriptaddress]{revtex4}
\usepackage[dvips]{color}
\usepackage{graphicx}
\usepackage{epsfig}
\usepackage{dcolumn}
\usepackage{bm}
\usepackage{amsmath}
\usepackage{amssymb}

\begin{document}
\title{Scalar and vector Keldysh models in the time domain}
\author{M.N. Kiselev}
\affiliation{International Center for Theoretical Physics, Strada
Costiera 11, 34014 Trieste, Italy}
\author{K. Kikoin}
\affiliation{School of Physics and Astronomy, Tel-Aviv University
69978, Israel}

\date{\today}

\begin{abstract}
The exactly solvable Keldysh model of disordered electron system in
a random scattering field with extremely long correlation length is
converted to the time-dependent model  with extremely long
relaxation. The dynamical problem is solved for the ensemble of
two-level systems (TLS) with fluctuating well depths having the
discrete $Z_2$ symmetry. It is shown also that the symmetric TLS
with fluctuating barrier transparency may be described in terms of
the planar Keldysh model with dime-dependent random planar rotations
in $xy$ plane having continuous $SO(2)$ symmetry. The case of
simultaneous fluctuations of the well depth and barrier transparency
is subject to non-abelian algebra. Application of this model to
description of dynamic fluctuations in quantum dots and optical
lattices is discussed.
\end{abstract}


\maketitle

The model with infinite  correlation range of fluctuating fields
$V(\bf r)$ proposed by L.V. Keldysh \cite{keld65} is one of few
exactly solvable problems in the theory of disordered electron
systems. The approximation
\begin{equation}\label{re1}
D({\bf r}-{\bf r}')=\langle V({\bf r})V({\bf r'})\rangle= W^2
\end{equation}
makes identical all diagrams for the electron Green function (GF) in
the order $V^{2n}$. As a result summation of diagrammatic series in
the "cross technique" \cite{AGD} reduces to the problem of
calculation of combinatoric coefficient $A_n$ (number of pairwise
coupling of scattering vertices). In fact $A_n=(2n-1)!!$ is the
total number of identical diagrams in the order $2n$. The
perturbation series is summed exactly \cite{keld65,efros70}, and one
deals with averaging of ensemble of samples with constant $V$ but
the magnitude of this field randomly changes from realization to
realization.  In momentum space the correlation function (\ref{re1})
transforms into $D({\bf q})=(2\pi)^3W^2\delta({\bf q})$. The
electron GF in Keldysh model is averaged with Gaussian distribution
function characterized by the variance $W^2$.

This model is not widely used in current literature because it is
difficult to propose an experimental device, where the conjecture
(\ref{re1}) could be realized (see, however, \cite{shbz}). In the
present paper we discuss a realization of Keldysh model in
{\textit{time domain}}. In this case the analog of infinite spatial
correlation is the long memory effect, which can be realized in many
physical situations (see below). The structure of perturbation
series in time-dependent Keldysh model (TDKM) is the same as in
original one, and the long characteristic times of dynamical
correlation play the same part as infinite range spatial correlation
of static random potentials. Since the time axis is the only
coordinate in this problem, its effective dimension is "0+1".
Moreover, the TDKM admits natural generalization of original Keldysh
model. We will show that the dynamical fluctuations in time domain
may be both of scalar and of vector character. The kinematic
constraint existing in the vector TDKM results in elimination of
essential part of diagrams in cross technique, but the summation of
perturbation series is still exact. It results in 2D Gaussian
averaging for the GF in dynamical random field.

Leaving for the last section the discussion of real systems, where
TDKM arises as a description of generic disorder, we start with a
toy model of an ensemble of non-interacting two-level systems (TLS)
in a randomly fluctuating environment. In standard realization of
TLS, namely a double-valley well, particles are distinguished not
only by conventional quantum numbers but also by their position in
the well characterized by the index $j=l,r$ of the left $(l)$ or
right $(r)$ valley. The barrier between the valleys is characterized
by the tunneling matrix element $\Delta_0$. The Hamiltonian of
isolated TLS has the form
\begin{equation}\label{re2}
H^{(0)}_{{\mbox {\tiny TLS}}}=\sum_j \left(\varepsilon_j n_j +
Un_j^2\right) - \Delta_0(c^\dag_l c^{}_r +{\rm H.c.}).
\end{equation}
Here $n_j=c^\dag_j c^{}_j$ is the particle occupation number,
$\varepsilon_j$ is the discrete energy level in the valley $j$ and
$U$ is the interaction parameter for two particles in the same
valley. The condition $U\gg \Delta$ is usually assumed.  We consider
spinless particles, having in mind that the theory can be applied
both to interacting bosons and fermions (electrons) with frozen spin
degrees of freedom. To be specific we discuss tunneling electrons as
an example.

We start with the singly occupied TLS, where the constraint
$N=\sum_i n_i=1$ is imposed on the Hamiltonian, introduce
pseudospin operators $\sigma^+ =c^\dag_l c^{}_r,\sigma^- =c^\dag_r
c^{}_l, 2\sigma_z = n_l - n_r$, and reduce (\ref{re2}) to
\begin{equation}\label{re3}
H^{(0)}_{{\mbox {\tiny TLS}}}= -\delta_0 \sigma_z - \Delta_0
\sigma_x -\mu_0 (N-1).
\end{equation}
in the pseudospin subspace. Here the asymmetry parameter
$\delta_0=\varepsilon_r-\varepsilon_l$ play the role of effective
"magnetic" field, the Lagrange parameter $\mu_0$ controls
constraint.

The scalar fluctuation field is introduced as random fluctuations of
TLS asymmetry, namely as a time dependent field $\delta_\rho
(t)$$=$$\delta_0$$+$$h_\rho(t)$ determined by its moments
\begin{eqnarray}\label{1.3}
\overline{h_\rho(t)}=0,\;\;\;\;\; \overline{h_\rho(t)h_\rho(t+\tau)}
= D(\tau).
\end{eqnarray}
Here the overline stands for the ensemble average. Thus we reduced
the original model to the effective spin Hamiltonian in magnetic
field with random time-dependent component. The problem
 can be reformulated as
a study of propagation of fermions along the time axis in the
presence of time-dependent random scalar potential
$\delta_{\rho}(t)(n_r-n_l)/2$, and the cross technique may be used
in calculation of the propagators \cite{KKAR}.

The analog of Keldysh conjecture in time domain is a slowly varying
random field $\sim$$\exp$$($$-$$\gamma$$t$$)$. A very long
relaxation time $\tau_{rel}$$\sim 1$$/$$\gamma$ with small $\gamma$
is presumed, so that the noise correlation function is given by
\begin{equation}
D(\omega)=\lim_{\gamma\to 0}
\frac{2\zeta^2\gamma}{\omega^2+\gamma^2}=2\pi
\zeta^2\delta(\omega) \label{fluct}
\end{equation}
(the noise correlation function (\ref{fluct}) is normalized  in
such a way that the corresponding vertices are dimensionless). In
this limit the averaged spin propagator describes the ensemble of
states with a field $\delta =const$ in a given state, but this
constant is random in each realization. The "planar-type" TDKM may
be derived in a similar way. For this sake one should introduce a
random component in the tunneling matrix element,
$\Delta=\Delta_0+\Delta_\rho(t)$ with
$\overline{\Delta_\rho(t)}=0$ and make similar conjecture
(\ref{fluct}) about the correlation function
$F(\tau)=\overline{\Delta_\rho (t)\Delta_\rho (t+\tau)}$, namely
approximate its Fourier transform by
\begin{equation}
F(\omega)=\lim_{\gamma\to 0}
\frac{4\xi^2\gamma}{\omega^2+\gamma^2}=4\pi \xi^2\delta(\omega)
\label{fluctvec}
\end{equation}

Thus, we treat our toy Hamiltonian (\ref{re3}) in the following way.
First we study the limiting case $\delta_0\gg \Delta_0$, were the
interdot tunneling is considered as a perturbation to the
\textit{longitudinal} term affected by stochastization in the scalar
TDKM. This problem may be solved by means of the standard technique
\cite{keld65,efros70,sad} generalized for the time-domain case. Then
we turn to the case of \textit{transversal} random tunneling
potential and solve this problem by means of correspondingly
modified "planar" TDKM.

We start with the scalar TDKM and treat the term
$H^{(0)}_\parallel=\delta_0(n_r-n_l)/2$ as a zero order
approximation with electron occupying the level $\varepsilon_l$ in
the ground state. Without stochastic perturbation the role of the
tunneling term $H^{}_\perp=-\Delta_0 \sigma_x$ is in admixing a
charge transfer exciton to the ground state with the corresponding
energy level shifts,  $\varepsilon_{l,r} \to \varepsilon_{l,r} \mp
\Delta^2_0/\delta_0$ for the ground and excited states,
respectively. The time-dependent perturbation stochastisizes this
simple picture.

\begin{figure}[h]
  \includegraphics[width=2cm,angle=0]{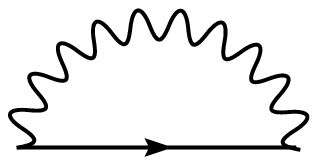}\hspace*{5mm}
  \includegraphics[width=2cm,angle=0]{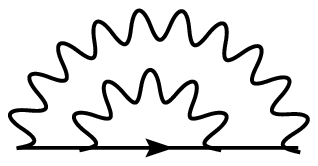}\hspace*{5mm}
  \includegraphics[width=2cm,angle=0]{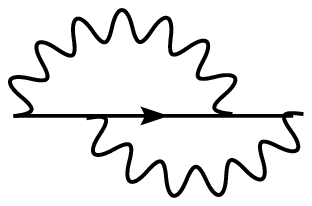}
  \caption{Irreducible Feynman diagrams for scalar TDQM.
  Solid and wavy lines stand for $g_i(\varepsilon)$ and
  $D(\omega)$.
  } \label{f.self1}
\end{figure}

Let us introduce the retarded propagators for the scalar TDKM
\begin{equation}\label{get}
G^R_{j,s}(t-t')=\langle
c{}_j(t)c^\dag_j(t')\rangle_R=-i\langle[c^{}_j(t)c^\dag_j(t')]_+\rangle.
\end{equation}
and consider their evolution on the time axis under the influence of
random component $h_j(t)$ ("random longitudinal magnetic field" in
pseudospin notation), first assuming $\gamma$$\to$$0$,
$\Delta_0$$/$$\gamma\to 0$. After averaging $G^R_{j}(t-t')$ in
accordance with (\ref{1.3}) and making the Fourier transformation by
means of (\ref{fluct}), we come to the series
\begin{equation}\label{seriess}
G^R_{j,s}(\varepsilon) = g_j(\varepsilon)\left[1+\sum_{n=1}^\infty
A_n \zeta^{2n}g_j^{2n}(\varepsilon)
 \right]
\end{equation}
Here $g_l(\varepsilon)=(\varepsilon+i\eta)^{-1}$ and
$g_r=(\varepsilon+\delta_0+i\eta)^{-1}$ are the bare propagators,
$A_n=(2n-1)!!$ is the above mentioned combinatoric coefficient (see
Fig. \ref{f.self1}, where several first irreducible diagrams are
shown).

Like in the real space Keldysh model \cite{keld65}, the series
(\ref{seriess}) may be summed by means of the integral
representation\cite{irp1} for $($$2$$n$$-$$1$$)$$!$$!$ . Then
changing the order of summation and integration (Borel summation),
one comes to the following equation for the left valley GF
\begin{equation}\label{GR}
G^R_{l,s}(\varepsilon)=\frac{1}{\zeta\sqrt{2\pi}}\int_{-\infty}^{\infty}
e^{-z^2/2\zeta^2}\frac{dz}{\varepsilon-z+i\eta}
\end{equation}
Remarkably, the single electron GF in this model has no poles,
singularities or branch cuts. Similar procedure may be applied to
the Green function $G^R_r$. As a result of this Gaussian averaging
the "magnetization" $\overline \sigma_z$ is reduced and the
corresponding response to transversal field is modified accordingly.

Next, we formulate the "planar" TDKM for the bare Hamiltonian
$H^{(0)}_{{\mbox {\tiny TLS}}}$ (\ref{re2}) with impenetrable
barrier $\Delta_0 \to 0$ and symmetric valleys,
$\varepsilon_l=\varepsilon_r$, isolated from each other. The
transverse random perturbation is introduced by
\begin{equation}\label{htrans2}
H_\rho(t)=\Delta_\rho(t)\sigma^+ +\Delta^*_\rho(t)\sigma^{-}
\end{equation}
so that the inter-valley tunneling is stochastisized by means of
averaging in time-domain with correlation function
(\ref{fluctvec}). The tunnel matrix element $\Delta$ is
transformed as $\Delta_\rho$$\to$$ \Delta_\rho
e^{i(\varphi_r-\varphi_l)}$ under the gauge transformation $c_j\to
c_je^{i\varphi_j}$, and we presume $\Delta_\rho$ to be  a complex
variable.

\begin{figure}[h]
\begin{center}
\includegraphics[width=3.5cm,angle=0]{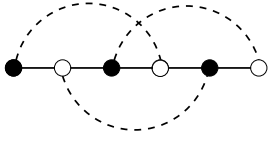}
\caption{First non-vanishing  vertex correction to the Green
function self-energies in the planar Keldysh model. Black and white
sites correspond to two terms in the Hamiltonian (\ref{htrans});
transversal pseudospin  correlation functions $F(\omega)$
(\ref{fluctvec}) are represented by dashed lines.}\label{f.5}
\end{center}
\end{figure}

Unlike the scalar TDKM, the noticeable part of diagrams in the
perturbation series disappears due to the kinematical restrictions
$\sigma^+\sigma^+=\sigma^-\sigma^-=0$ (see also \cite{com1}). Only
the diagrams with pseudospin operators ordered as $\ldots
\sigma^+\sigma^-\sigma^+\sigma^-\ldots $ survive in the expansion
for the  GF of planar model
\begin{equation}\label{seriest}
G^R_{j,p}(\varepsilon) = g_j(\varepsilon) + \sum_{n=1}^\infty B_n
(\sqrt{2}\xi)^{2n}g_j^{2n+1}(\varepsilon).
\end{equation}
The vertices in the cross technique are now "colored" in
accordance with two terms entering the random potential. The
vertices with different colors have to be ordered in alternating
way, and the correlation lines connect only the vertices of
opposite color (see Fig. \ref{f.5}.)

As a result of the above kinematic restrictions, the combinatoric
coefficient $B_n=n!$. Then  we use the integral
representation\cite{irp2} for $n!$ and transform the series
(\ref{seriest}) into
\begin{eqnarray}
G^R_{j,p}(\epsilon)=g_i(\epsilon)\left\{
    1+2\sum_n\int_0^\infty t dt
    \left[t\sqrt{2}\xi g_j(\epsilon)]^{2n}
    \right]e^{-t^2}
\right\}\nonumber
\end{eqnarray}
Here we substituted $t^2$ for the variable $z$. Then changing the
order of summation and integration, we transform $G^R_j(\epsilon)$
into the integral
\begin{equation}\label{a2.int1}
G^R_{j,p}(\epsilon)= \int_0^\infty
    2tdt\frac{g_j(\epsilon)}{1-2t^2\xi^2g_j^2(\epsilon)}e^{-t^2}
\end{equation}
Taking into account the explicit form of the free
 propagator $g_j(\epsilon)$, we change the
integration variable once more, $t=u/\sqrt{2}\xi$, and transform
(\ref{a2.int1}) into

\begin{equation}\label{a2.int2}
G^R_{j,p}(\epsilon)=\int_0^\infty \frac{udu}{2\xi^2}\left(
    \frac{1}{\epsilon -u+i\eta}+ \frac{1}{\epsilon+u+i\eta}\right)
    e^{-u^2/2\xi^2}
\end{equation}
Now we introduce the "cartesian" coordinates, $x=u\cos \phi,
y=u\sin\phi$, so that $u=\sqrt{x^2+y^2}$ and $dxdy=udud\phi$. The
angle independent integral (\ref{a2.int2}) may be rewritten as
\begin{eqnarray}\label{fvect}
&& G^R_{j,p}(\epsilon)=\frac{1}{2}\int_{-\infty}^{+\infty}
\frac{dxe^{-x^2/2\xi^2}}{\xi\sqrt{2\pi}} \int_{-\infty}^{+\infty}
\frac{dye^{-y^2/2\xi^2}}{\xi\sqrt{2\pi}}\nonumber \\
&&
    \left[
\frac{1}{\epsilon-\sqrt{x^2+y^2}+i\eta}+\frac{1}{\epsilon+\sqrt{x^2+y^2}+i\eta}
    \right].
\end{eqnarray}
This result is a natural generalization of the one-dimensional
Gaussian averaging (\ref{GR}) characteristic for the scalar TDKM to
the two-dimensional Gaussian averaging of planar random field with
purely transversal $(xy)$ fluctuations. Only the modulus of random
field $r=\sqrt{x^2+y^2}$ is averaged, whereas the angular variable
remains irrelevant due to the in-plane isotropy of the system. Like
in the scalar model, the averaged GF has no singularities.

Although the GF lost its pole structure, the standard Feynman rules
for construction of irreducible parts (Figs. \ref{f.self1},
\ref{f.5}) and corresponding skeleton diagrams \cite{AGD,efros70}
are still valid. However, the important reservation should be made:
the self energy  cannot be treated as a renormalization of bare pole
because the bare and dressed GF are connected by non-local integral
operators [see Eqs. (\ref{GR}),(\ref{a2.int2})].
 Nevertheless, the ordinary differential equation
connecting the GF and its derivative over energy can be obtained for
both versions of TDKM. This equation was found for the scalar
Keldysh model in \cite{efros70}. Here we derive this equation
without appealing to the Ward identity and then generalize the
derivation procedure for the planar model.

To calculate the derivative $dG(\varepsilon)/d\varepsilon$ for the
scalar TDKM, we start with expansion (\ref{seriess}) (index $s$ is
omitted below for the sake of brevity). It is convenient to count
the energy off the position of chemical potential $\mu_0$ [see Eq.
(\ref{re3})] in the middle between the levels $\varepsilon_{l,r}$.
So, we shift the energies $\varepsilon \to \varepsilon_{\pm}=
\varepsilon \mp \delta_0/2$ for the left and right GR, respectively.

The same procedure, which leads to Eq. (\ref{GR}) for GR gives the
following equation for its derivative:
\begin{equation}\nonumber
\frac{dG}{d\varepsilon}=-\frac{g^2(\varepsilon_\alpha)}{\zeta\sqrt{2\pi}}\int_{
-\infty}^\infty e^{-\displaystyle
\frac{z^2}{2\zeta^2}}\left[1+\frac{1}{\zeta^2}
\frac{z^3}{\varepsilon_\alpha-z+i\eta}\right]dz.
\end{equation}
($\alpha=\mp$). Calculating the integral and substituting
$g^2(\varepsilon_\alpha)=\varepsilon_\alpha^{-2}$ we come to the
differential equation
\begin{equation}\label{difscal}
\zeta^2\frac{dG(\varepsilon_\alpha)}{d\varepsilon}=1
-\varepsilon_\alpha G(\varepsilon_\alpha)
\end{equation}
similar to that obtained for the real space scalar Keldysh model
\cite{efros70,sad}.

\begin{figure}[h]
\begin{center}
  \includegraphics[width=1.5cm,angle=0]{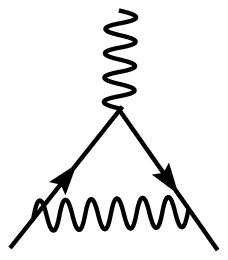}\hspace*{10mm}
  \includegraphics[width=1.5cm,angle=0]{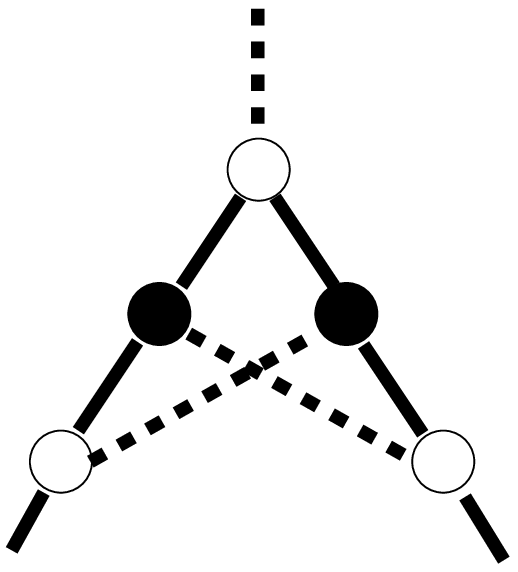}
  \end{center}
  \caption{First non-vanishing vertex diagrams for the   $
  \Gamma$ in
  scalar (left) and planar (right) TDKM. }\label{f.vert1}
\end{figure}

Generalization of this procedure for the symmetric planar model
($\delta_0=0$) is more cumbersome. We start with differentiating the
series (\ref{seriest}) over the energy. Then the analog of Eq.
(\ref{a2.int1}) for the derivative has the form
$$
\frac{dG}{d\varepsilon}= -g^2(\varepsilon)\left[1+\int_0^\infty 2tdt
\frac{2(2t^2-1)t^2\xi^2
g^2(\varepsilon)}{1-2t^2\xi^2g^2(\varepsilon)} e^{-t^2}\right]
$$
The subsequent variable change which gave Eq. (\ref{a2.int2}) for
the GF gives for its derivative the following equation
\begin{equation}\label{deriv}
\frac{dG}{d\varepsilon}= -g^2(\varepsilon)\left[1+
\frac{1}{2}\left(\frac{J_4}{\xi^4}-\frac{J_2}{\xi^2} \right)\right]
\end{equation}
where
$$
J_n =\int_0^\infty dz z^n
\exp\left(-\frac{z^2}{2\xi^2}\right)\left[g(\varepsilon-z)+(-1)^{n+1}
g(\varepsilon+z) \right]
$$
After some manipulations, these integrals are represented via the GF
for the planar model (\ref{a2.int2}):
\begin{equation}
J_2 = 2\varepsilon \xi^2 G -2\xi^2,~~ J_4= -4\xi^4+ \varepsilon^2
J_2
\end{equation}
Substituting these integrals in Eq. (\ref{deriv}), we come
eventually to the differential equation
\begin{equation}\label{difvect}
\xi^2 \frac{dG(\varepsilon)}{d\varepsilon} =1-\varepsilon
G(\varepsilon)\left(1-\frac{\xi^2}{\varepsilon^2}\right)
\end{equation}
which is obviously the generalization of Eq. (\ref{difscal}). These
two equation may be rewritten in a unified way:
\begin{eqnarray}\label{dif}
\varepsilon_\alpha G_{\alpha, s} -1 &=&
\zeta^2G_{\alpha, s}^2\frac{d}{d\varepsilon}G_{\alpha, s}^{-1} \nonumber \\
\varepsilon G_p -1 &=& \xi^2
G_p^2\left[\frac{1}{\varepsilon}\frac{d}{d\varepsilon}\left(
\varepsilon G_p^{-1}\right) \right]
\end{eqnarray}
The solutions of these equations satisfying the boundary condition
$G(\epsilon\to \infty)=\epsilon^{-1}$ is given by (\ref{GR},
\ref{a2.int2}). It is worth noting that the differential operator in
the right hand side of the second equation is nothing but ${\rm
div}_\varepsilon$ in polar coordinates. This form reflects effective
two-dimensionality of Gaussian averaging in the planar TDKM. Now one
may introduce vertex parts using the analogy with the Ward
identities
\begin{eqnarray}\label{ward}
   \Gamma_s= \frac{d}{d\varepsilon} G_s^{-1},~~~
   \Gamma_p= \frac{1}{\varepsilon}\frac{d}{d\varepsilon}
   (\varepsilon
   G_p^{-1}).
\end{eqnarray}
These vertices,  together with equations (\ref{dif}) will be useful
for calculation of response functions of our TLS (see below). Like
in the self energy parts (Figs. \ref{f.self1}, \ref{f.5}), the
planar TDKM lacks most of diagrams of scalar model due to the
kinematic restrictions: the sites in the vertices of triangle are of
the same color, black and white sites alternate, and dashed lines
connect sites of opposite colors. First nonvanishing vertices for
both models are shown in Fig. \ref{f.vert1}

\begin{figure}[h]
\vspace*{-5mm}
\begin{center}
  \includegraphics[height=3.5cm,width=4.5cm,angle=0]{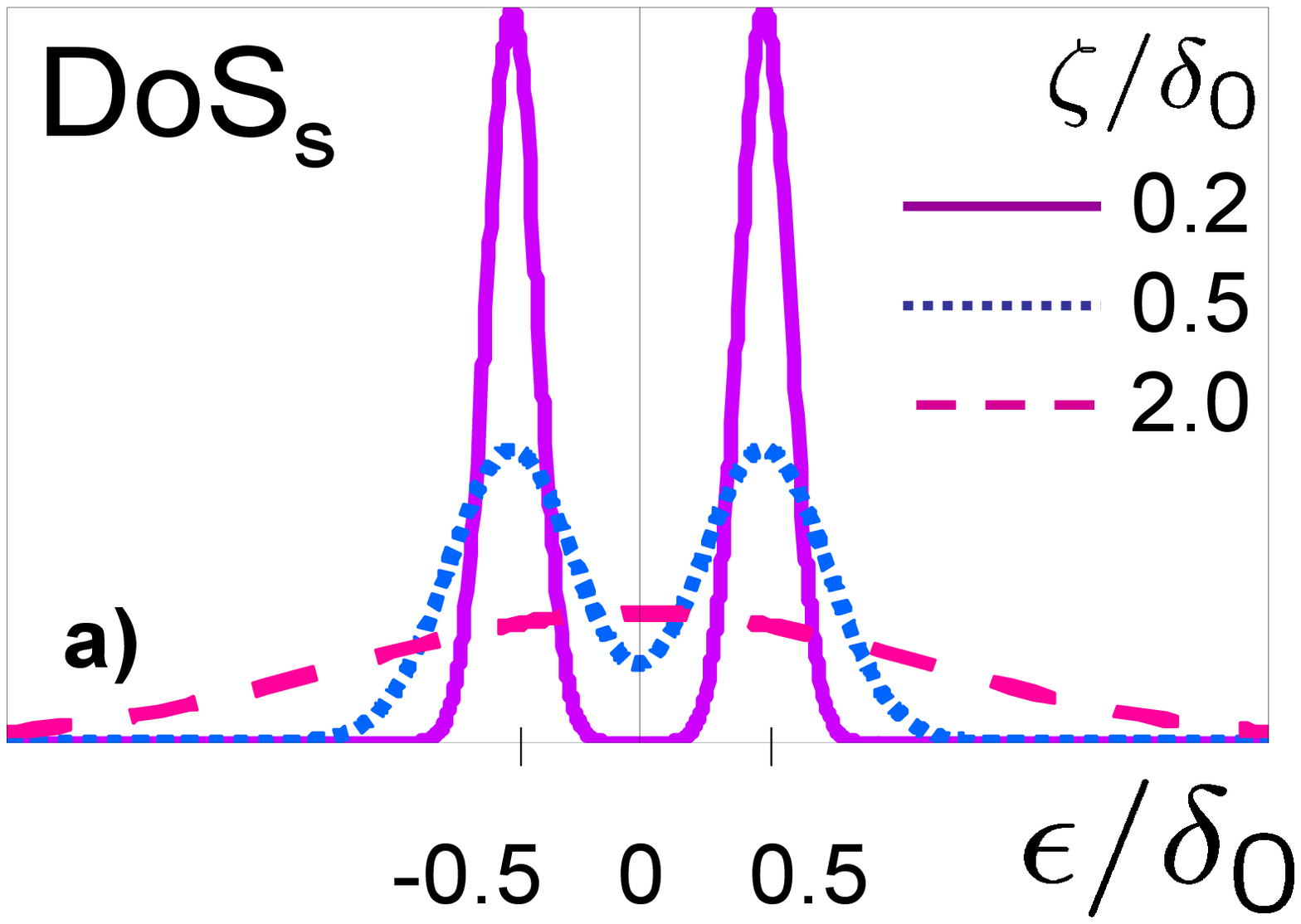}
  \hspace*{-1cm}
  \includegraphics[height=3.5cm,width=4.5cm,angle=0]{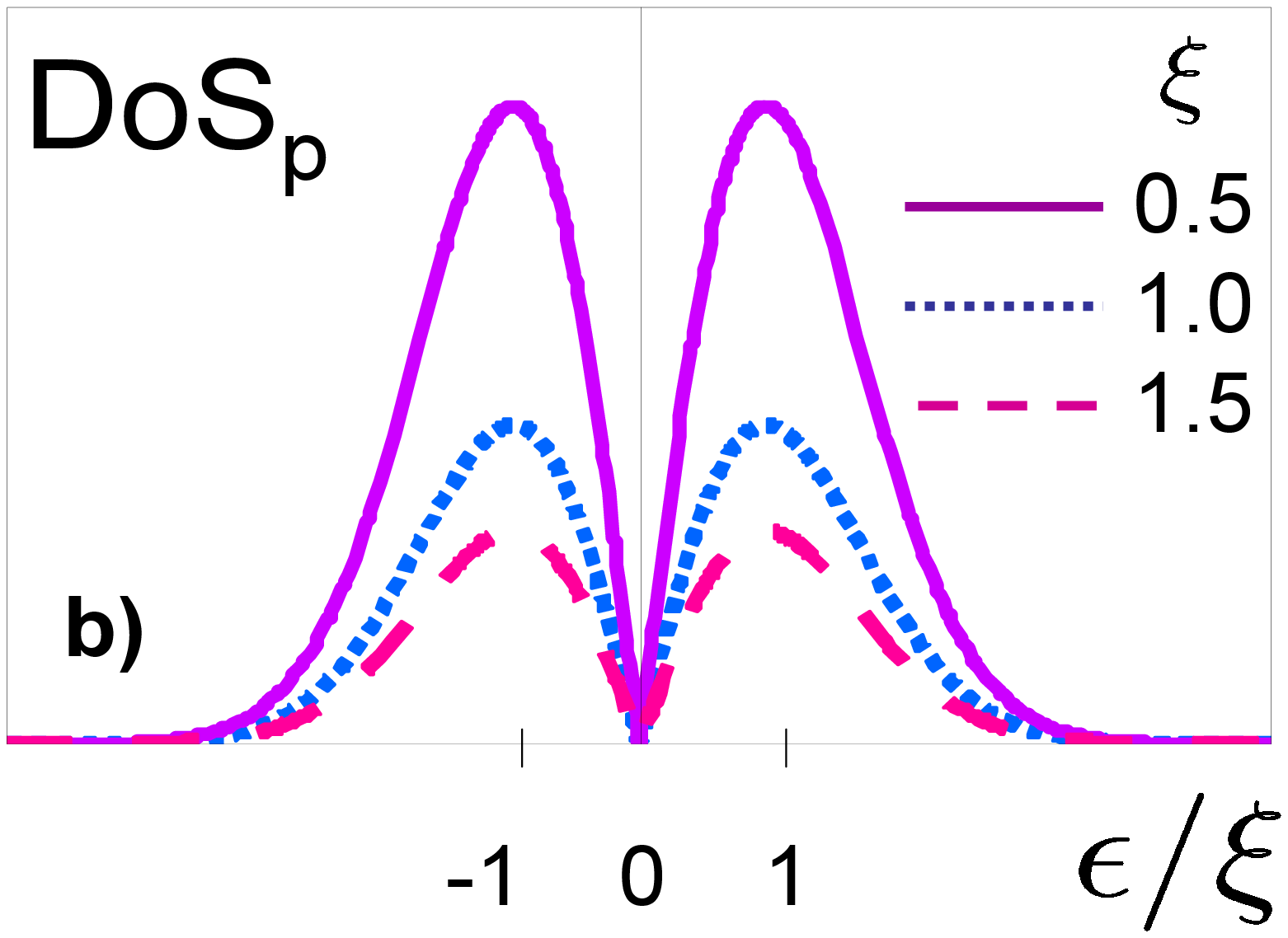}
\end{center}
\vspace*{-7mm}
  \caption{Density of states in scalar (left) and planar
  (right) model (all units are arbitrary).}\label{f.dos}
\end{figure}

The density of states (DoS) in stochastisized TLS is given by the
imaginary parts of GF (\ref{GR}), (\ref{a2.int2}) for scalar and
planar TDKM, respectively:
\begin{eqnarray}\label{doss}
\nu_s(\varepsilon) = \frac{2}{\zeta\sqrt{2\pi}}
\exp\left(-\frac{\varepsilon^2+\delta_0^2}{2\zeta^2}\right)
\cosh\left(\frac{\varepsilon\delta_0}{\zeta^2}\right),
\end{eqnarray}
\vspace*{-5mm}
\begin{eqnarray}\label{dosv}
\nu_p(\varepsilon) = \frac{1}{\xi^2}|\varepsilon|
\exp\left(-\frac{\varepsilon^2}{2\xi^2}\right).
\end{eqnarray}
In the scalar model $\nu_s(\varepsilon)$ is a superposition of two
Gaussians centered around $\varepsilon_l$ and $\varepsilon_r$,
respectively. In the planar model $\nu_p(\varepsilon)$ is
represented by a single Gaussian with a dip "burnt" around zero
energy (Fig. \ref{f.dos}).

Switching on the tunneling term $H_\perp$ in scalar model without
random field, we come to the picture of two levels mutually repulsed
due to coherent interdot tunneling and broadened due to incoherent
time-dependent intradot fluctuations. The spectrum is still gapful
at small enough ratio $\zeta/\delta_0$, If the $\zeta \lesssim
\delta_0$, the DoS merges into double hump Gaussian structure. The
information about position of electron in right or left valley is
completely lost at $\zeta/\delta_0 > 1$. In the planar model for
symmetric TLS instead two-peak structure due to avoided crossing
characteristic for coherent tunneling, we get a pseudogap around
zero energy due to stochastic tunneling, which survives at any
variance $\xi$.

Let us allow now the fluctuations of both longitudinal and
transverse components which corresponds to simultaneous
fluctuations of the well depth and barrier transparency for the
case of symmetric TLS ($\delta_0=0$). We solve this problem by
means of path integral formalism. The Lagrangian and corresponding
action are defined on the Keldysh contour $K$ (see e.g.
\cite{Keldysh_t})
\begin{eqnarray}
{\cal L}(t)=\sum_{j=l,r}\bar c_j i\partial_t c_j -
H,\;\;\;\;\;S_K=\int_K{\cal L}(t)d t.
\end{eqnarray}
Here  $\bar c_j, c_j$ are Grassmann variables describing the
electron.  The time-dependent gauge transformation $c_j(t)\to
c_j(t) e^{i\varphi_j(t)}$,  converts the fluctuation of the well
depth to the fluctuation of the {\it phase} of the tunnel matrix
element under the choice
\begin{eqnarray}
\varphi_j(t)=\int_{-\infty}^t h_j(t') dt'.
\end{eqnarray}
We therefore identify the longitudinal and transverse noise with
phase fluctuations of the barrier transparency and fluctuations of
the modulus of the tunnel matrix element, respectively and unify
them in the path integral description.

The ensemble averaging
$$
\langle ... \rangle_{noise}=\int dh_\rho P_{l}(h_\rho)\int
d\Delta^*_\rho d\Delta_\rho P_{tr}(\Delta^*_\rho,\Delta_\rho) ...
$$
is done with the help of probability distribution functions for
longitudinal and transverse fluctuations
$$
P_{l}=\frac{1}{\zeta\sqrt{2\pi}}\exp{\left(-\int_K d t d
t'h_\rho(t)D^{-1}(t-t')h_\rho(t')\right)}
$$
$$
P_{tr}=\frac{1}{2\pi \xi^2}\exp{\left(-\int_K d t d
t'\Delta^*_\rho(t)F^{-1}(t-t')\Delta_\rho(t')\right)}.
$$
The GF's can be calculated by means of generating functional
corresponding to Keldysh action $S_K$ in a standard way
\cite{Popov}, \cite{NOR}. In the "infinite memory" limit
(\ref{fluct}), (\ref{fluctvec}) we easily express the GF of the
electron in the symmetric double well potential
\begin{eqnarray}
G^R_{j,v}(\epsilon)=\frac{1}{\zeta\xi^2(2\pi)^{3/2}}\int_{-\infty}^\infty
dz e^{-z^2/2\zeta^2} \nonumber
\end{eqnarray}\vspace*{-6mm}
\begin{eqnarray}\label{gr3a}
\int dw^* dw e^{-|w|^2/2\xi^2} \frac{\epsilon \pm z}{(\epsilon
+i\eta)^2-z^2-|w|^2}
\end{eqnarray}
Noticing that the GFs do not depend on the well index $j$ and
performing the integration over angles in spherical coordinate
system we get
\begin{eqnarray}
G^R_{v}(\epsilon)=\frac{1}{2\xi}\int_0^{\infty}d\rho \rho
\exp\left(-\frac{\rho^2}{2\xi^2}\right) \nonumber
\end{eqnarray}\vspace*{-6mm}
\begin{eqnarray}\label{gr3b}
\frac{{\rm
erf}\left(\rho\sqrt{\frac{\xi^2-\zeta^2}{2\xi^2\zeta^2}}\right)}{\sqrt{\xi^2-\zeta^2}}
\left(
    \frac{1}{\epsilon -\rho+i\eta}+
    \frac{1}{\epsilon+\rho+i\eta}\right).
\end{eqnarray}
The Eqs (\ref{gr3a}, \ref{gr3b}) generalize  (\ref{GR}) and
(\ref{fvect}) for the case of anisotropic vector Keldysh model. The
three-dimensional Gaussian averaging in (\ref{gr3a}) stands for
vector character of the random field distributed on an ellipsoid.
Typical size of semi-axes is defined by the variances of
longitudinal and transverse noises. Like in the scalar an planar
models, the averaged GF has no singularities. The angle $\phi$
dependence is absent in (\ref{gr3a}) due to the in-plane isotropy of
the model $P_{tr}=P_{tr}(|\Delta|^2)$ which is preserved here. The
limits of strong easy axis $\xi$$\to$$0$ and easy plane
$\zeta$$\to$$0$ anisotropy correspond to scalar and planar models,
correspondingly.

We notice that the DoS for vector TDKM is also characterized by a
pseudogap. The energy dependence $\nu_v(\varepsilon)$$\sim$$
\varepsilon^2$ for small energies should be contrasted with
$\nu_s(\varepsilon)$$\sim$$const$ for scalar
$\nu_p(\varepsilon)$$\sim$$\varepsilon$ for planar model behavior.
This dependence reflects the fact that the probability to remain
close to initial level position decreases with increase of effective
dimensionality of the problem due to stochastization of the complex
tunneling matrix element.

One should specially mention the degenerate case of the isotropic
vector TDKM, characterized by $\xi$$=$$\zeta$$=$$\lambda$ and
describing rotation of pseudospin on a Bloch sphere. Performing
simple algebra we get an equation for GF
\begin{eqnarray}
\varepsilon G_v -1 = \lambda^2
G_v^2\left[\frac{1}{\varepsilon^2}\frac{d}{d\varepsilon}\left(
\varepsilon^2 G_v^{-1}\right) \right]
\end{eqnarray}
and the Ward Identity
\begin{eqnarray}
\Gamma_v= \frac{1}{\varepsilon^2}\frac{d}{d\varepsilon}
   (\varepsilon^2
   G_v^{-1}).
\end{eqnarray}
The DoS for the isotropic model is given by the expression
\begin{eqnarray}
\nu_v(\varepsilon) = \frac{1}{\lambda^3\sqrt{2\pi}}\varepsilon^2
\exp\left(-\frac{\varepsilon^2}{2\lambda^2}\right).
\end{eqnarray}

\begin{figure}[h]
\begin{center}
  \includegraphics[width=1.6cm,angle=0]{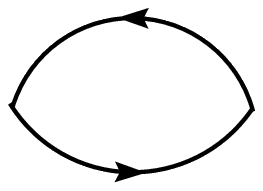}\hspace*{5mm}
  \includegraphics[width=1.6cm,angle=0]{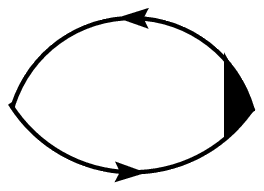}
\end{center}
\vspace*{-3mm}
  \caption{Diagrams for bare loop (transverse susceptibility) and
  loop with dressed vertex (longitudinal susceptibility).}\label{f.loop1}
\end{figure}

Our next task is calculation of response functions. In scalar TDKM
the longitudinal susceptibility is given by the correlation
function $\chi_\parallel(\omega)=i\int dt \exp{(i\omega t)}\langle
\sigma_z(t)\sigma_z(0)\rangle_R$, which is represented by two
loops with vertex corrections (Fig. \ref{f.loop1}). Here both
solid lines correspond either to $j=l$ or to $j=r$.
We confine ourselves with calculation of static susceptibility,
$\omega \to 0$. In order to work at finite temperatures we turn to
Matsubara Green functions functions ${\cal G}(i\epsilon_n)$ (similar
calculation can be done on the Keldysh contour in the real time path
integral formalism) and susceptibility
\begin{eqnarray}\label{chi}
\chi(i\omega_m)=
 T\sum_{n,j}{\cal G}_j(i\omega_m+i\epsilon_n){\cal
G}_j(i\epsilon_n)\Gamma(i\epsilon_n, i\omega_m)
\end{eqnarray}

 Then the first of Ward identities (\ref{ward})
provides us with the exact equation for the vertex
\begin{eqnarray}
\zeta^2(G_{j,s}^R)^2\Gamma_{j,s}^R(\epsilon, 0)=\epsilon
G_{j,s}^R-1, \label{GamR}
\end{eqnarray}
giving an access to the exact evaluation of $\chi(0)$. Combining
(\ref{chi}) with (\ref{GamR}) and analytical continuation of
(\ref{GR}), we find
\begin{eqnarray}\label{ssuca}
\chi_\parallel(0)=
-\sum_{\alpha=\mp}\int_{-\infty}^{\infty}\frac{ydy e^{-y^2/2}
}{\sqrt{2\pi}\zeta}n_F\left(\frac{(2y-\alpha\delta_0)\zeta}{2T}\right)
\end{eqnarray}
Here $n_F(x)$ is the Fermi distribution function. The asymptotic
behavior of static susceptibility $\chi(0)$ is
\begin{eqnarray}
\chi(0)\sim \left\{
\begin{array}{c}
1/T,\;\;\; T\gg (\zeta, \delta_0)\\
    1/\zeta,\;\;\; \zeta\gg (T,\delta_0) \label{Gm}
\end{array}\right.
\end{eqnarray}
There is no vertex correction to transverse susceptibility
$\chi_\perp(0)=\langle\sigma_+,\sigma_-\rangle_R$. It is given by
the bare loop formed by the left and right GF. In case of symmetric
TQD ($\delta_0=0$) and big variance $\zeta\gg \Delta_0$ this
function is as smooth as $\chi_\parallel(0)$ with changing $\zeta$
and $T$. Its asymptotic behavior is given by the same Eq.
(\ref{Gm}). The physical sense of these results in obvious: the
information about position of electron in a given well is lost at
strong enough stochastization $\zeta \gg T$.

Next, we calculate the susceptibility for planar model in case of
symmetric TLS with $\delta_0=0$. In this case $\chi_\perp(0)$
differs from  $\chi_\parallel(0)$ by factor 2, so that it is enough
to calculate the latter one. Now we appeal to the second equation
from (\ref{dif}) with $\Gamma_p$ defined in (\ref{ward}). Then like
in previous case the calculation of $\chi_\parallel(0)$ is reduced
to finding the combination $i\epsilon_n {\cal G}_p(i\epsilon_n)-1$.
Straightforward computation gives
\begin{eqnarray}\label{ssucv}
\chi_\parallel(0)=\frac{1}{\xi}\int_{0}^{\infty}y^2dy e^{-y^2/2
}\tanh\left(\frac{y\xi}{2T}\right).
\end{eqnarray}
The behavior of $\chi_\parallel$ as a function of $T$ and $\xi$ is
close to that for scalar model, including the asymptotic dependence
(\ref{Gm}). The equations for static susceptibilities in isotropic
vector TDKM can be easily found with help of (\ref{gr3a},
\ref{gr3b})
\begin{eqnarray}
\chi(0)=\frac{1}{\lambda\sqrt{2\pi}}\int_{0}^{\infty}y^3dy e^{-y^2/2
}\tanh\left(\frac{y\lambda}{2T}\right).
\end{eqnarray}
These susceptibilities are characterized by the same asymptotic
behavior as those for scalar/planar models.
\smallskip

The toy model of noninteracting TLS under dynamical stochastization
demonstrate some generic properties of TDKM: (i) the loss of
characteristic spin or pseudospin behavior at variance exceeding
temperature; (ii) the effective two-dimensionality of Gaussian
averaging in planar TDKM as its main distinction from scalar model;
(iii) the effective three-dimensionality of vector TDKM. These
features survive also in more realistic situations. One of possible
applications of this theory is the problem of electron tunneling
through double quantum dot in a regime of strong Coulomb blockade,
where the source of stochastization is a random time-dependent gate
voltage applied to one of the valleys \cite{KKAR}. The case of $N=2$
was considered, where the starting Hamiltonian $H^{(0)}$ is that of
Eq. (\ref{re2}) with added spin index. In this case the scalar TDKM
may be used in the limit of slow fluctuations (\ref{fluct}), the
double quantum dot looses its spin characteristics at low $T$, and
the Kondo-type zero bias anomaly is smeared accordingly \cite{KKR}.
The important difference between spinful and spinless TLS models is
in their symmetry. The symmetry of TLS with $N$$=$$2$ considered in
\cite{KKR} is $SO(5)$, which is reduced to $SO(3)$ for low-energy
part of excitation spectrum, so the Lie algebra is non-abelian.
However, it was possible to introduce scalar TDKM due to abelian
character of time-dependent random gauge field.

The symmetry of the scalar TLS with $N$$=$$1$ is given by the
discrete group $Z_2$ with abelian algebra. The gauge transformation
allows to identify the fluctuations of the scalar model as $U(1)$
fluctuations of the phase of tunnel matrix element. In the planar
model one deals only with the planar $(xy)$ rotations, so the
relevant symmetry is $SO(2)$ with still abelian algebra. The phase
fluctuation in that case are discrete $Z_2$ ($\phi$$=$$0$$,$$\pi$ to
provide the condition $\bar\Delta_\rho$$=$$0$), while modulus
fluctuations determine effective 2d behavior. The symmetry of
isotropic vector TDKM is $SU(2)$ and corresponding algebra is
non-abelian.

One may mention several more physical systems, where the scalar
and/or planar TDQM is useful. One of such models is the big quantum
dot with charge fluctuations accompanied by longitudinal and
transversal spin fluctuations \cite{KisGef}. The class of Gaussian
ensembles corresponds to infinite-range correlations in the charge
and spin sectors of the model. Both spin and charge interactions
contain stochastic component \cite{KAA}, leaving a room for original
formulation of the Keldysh model. The gauge field theory, based on
functional bosonization being formulated in the time-domain, opens a
possibility of stochastic treatment of dynamic processes. As is
shown in \cite{KisGef}, the transverse spin correlation function for
anisotropic spin exchange contains both short-time and long-time
correlation parts. While the short-time (white noise) correlations
dominate away from the Stoner instability, the (infinitely)
long-time correlations become important  as one approaches the
regime of strong fluctuations of the magnetization. The long-time
part of the model is equivalent to the planar TDKM.

Another interesting object is the optical superlattice consisting
of biased double wells \cite{Bloch}. The bias is random, but the
number of atoms in the same in all TLS in this experimental setup
due to the "interaction blockade". One may expect that the well
population in these TLS could be stochastized in accordance with
Fig. \ref{f.dos}, provided the Keldysh-type fluctuations
(\ref{fluct}) or (\ref{fluctvec}) with long relaxation times were
realized experimentally.

We thank Boris Altshuler, Jan von Delft, Yuri Galperin, Yuval Gefen,
Jean Richert and Wilhelm Zwerger for helpful discussions.


\begin{thebibliography}{99}
\bibitem{keld65} L.\,V. Keldysh, "Semiconductors in strong electric
field", D. Sci. Thesis, Lebedev Institute, Moscow, 1965.

\bibitem{AGD}
A.\,A. Abrikosov, L.\,P. Gorkov, and I.\,E. Dzyaloshinski, {\it
Methods of Quantum Field Theory in Statistical Physics},
Prentice-Hall, Englewood Cliffs, NJ, 1963.

\bibitem{efros70} A.\,L. Efros, Sov. Phys. -- JETP {\bf32}, 479 (1971)
[Zh. Eksp. Teor. Fiz. {\bf59}, 880 (1970)].

\bibitem{shbz} M.E.Raikh and T.V.Shahbazyan, Phys. Rev. {\bf B
47}, 1522 (1993).

\bibitem{KKAR} M.\,N. Kiselev, K. Kikoin, Y. Avishai, and J.
Richert, Phys. Rev. B {\bf74}, 115306 (2006).

\bibitem{sad} M.\,V. Sadovskii, {\it "Diagrammatica"}, World Scientific, Singapore,
2005.

\bibitem{irp1}$
(2n-1)!!=2^n\frac{1}{\sqrt{\pi}}\Gamma(n+1/2)=\frac{1}{\sqrt{2\pi}}\int_{-\infty}^\infty
dtt^{2n}e^{-\frac{t^2}{2}}. $

\bibitem{KKR} M.\,N. Kiselev, K. Kikoin, and J. Richert,
         arXiv:0803.2676;  Physica Status Solidi (c) ({\it in press})

\bibitem{com1}The
correlators $\overline{\Delta(t)\Delta(t+\tau)}$ and
$\overline{\Delta^*(t)\Delta^*(t+\tau)}$ are transformed under
"local" time-independent  gauge transformation and therefore average
to zero according to Elitzur theorem [S. Elitzur, Phys. Rev. {\bf D
12}, 3978 (1975)].

\bibitem{irp2}$n!=\Gamma(n+1)=\int_0^\infty dz z^ne^{-z}.$

\bibitem{Keldysh_t} L. V. Keldysh, Sov. Phys. -- JETP {\bf20}, 1018 (1965)
[Zh. Eksp. Teor. Fiz. {\bf47}, 1515 (1964)].

\bibitem{Popov} V.N.Popov, {\it "Functional Integrals in Quantum
Field Theory and Statistical Physics"}, D. Reidel Publishing
Company, Dordrecht, 1983.

\bibitem{NOR} J.W. Negele and H. Orland, {\it "Quantum Many-Particle
Systems"}, Addison-Wesley, Reading, MA, 1988.

\bibitem{KisGef} M.\,N. Kiselev and Y. Gefen, Phys. Rev. Lett.
{\bf 96}, 066805 (2006).

\bibitem{KAA} I.\,L. Kurland, I.\,L. Aleiner, and B.\,L. Altshuler, Phys.
Rev. {\bf B 62}, 14886 (2000).

\bibitem{Bloch} P. Cheinet, S. Trotzky, M. Feld, U. Schnorrberger, M.
Moreno-Cardoner, S. F\"olling, and I. Bloch, Phys. Rev. Lett. {\bf
101}, 090404 (2008).

\end{thebibliography}
\end{document}